\documentclass{aa}

\newcommand{\hi}{H\,{\small{\sc I}}}
\newcommand{\hii}{H\,{\small{\sc II}}}

\newcommand{\msol}{$M_{\odot}$}

\usepackage{epsfig}
\usepackage{graphics}

\begin{document}

\hyphenation{con-sti-tu-ents there-by mar-gin-al-ly stat-is-tics}


\title{The mysterious H\,I deficiency of NGC\,3175}

\author{Michael Dahlem\inst{1}
\and 
Matthias Ehle\inst{2,3}
\and
Stuart D. Ryder\inst{4}
}

\institute{Sterrewacht Leiden, Postbus 9513, 2300 RA Leiden, The
  Netherlands
\and
XMM-Newton Science Operations Centre, Apartado 50727, 28080 
  Madrid, Spain
\and
Astrophysics Division, Space Science Department of ESA, ESTEC,
  2200 AG Noordwijk, The Netherlands
\and
Anglo-Australian Observatory, P. O. Box 296, Epping, NSW 1710,
Australia
}

\offprints{M. Dahlem}


\date{Received 6 June 2000 / Accepted 28 February 2001}

\abstract{
Australia Telescope Compact Array \hi\ observations reveal the 
existence of $5.8\,10^8$ \msol\ of \hi\ gas in the central 7 
kpc of the edge-on spiral galaxy NGC\,3175. The detected \hi\
and CO gas can explain why star formation, as traced by other 
emission processes, is going on in the inner part of its disk. 
On the other hand, the entire outer disk, beyond 3.5 kpc radius,
shows no \hi\ emission, has a very red colour and exhibits
neither radio continuum nor H$\alpha$ emission. This indicates 
that the outer part of NGC\,3175 is quiescent, i.e. not forming 
stars at a measurable rate.
Its \hi\ deficiency and the small extent of the \hi\ layer,
which is confined to the boundaries of the optically visible 
disk, make NGC\,3175 a peculiar spiral galaxy. 
No intergalactic \hi\ gas in the NGC\,3175 group was detected 
in our interferometric observations. Earlier Parkes telescope 
single dish \hi\ observations put an upper limit on the amount 
of diffuse gas that might have been missed by the interferometer 
at $2\,10^8$ \msol. On DSS plates no galaxy in the NGC\,3175 
group of galaxies (Garc\' \i a 1993) is close enough to it 
and none exhibits disturbances that could indicate a close 
interaction which might have led to the stripping of large
parts of its \hi\ gas. Thus, despite an extensive multi-wavelength 
investigation, the reason for the unusual absence of \hi\ and 
star formation activity in the outer disk of NGC\,3175 remains 
an intriguing mystery.
}

\maketitle

\section{Introduction}

\begin{figure*}[ht!]
\resizebox{0.9\hsize}{!}{\includegraphics{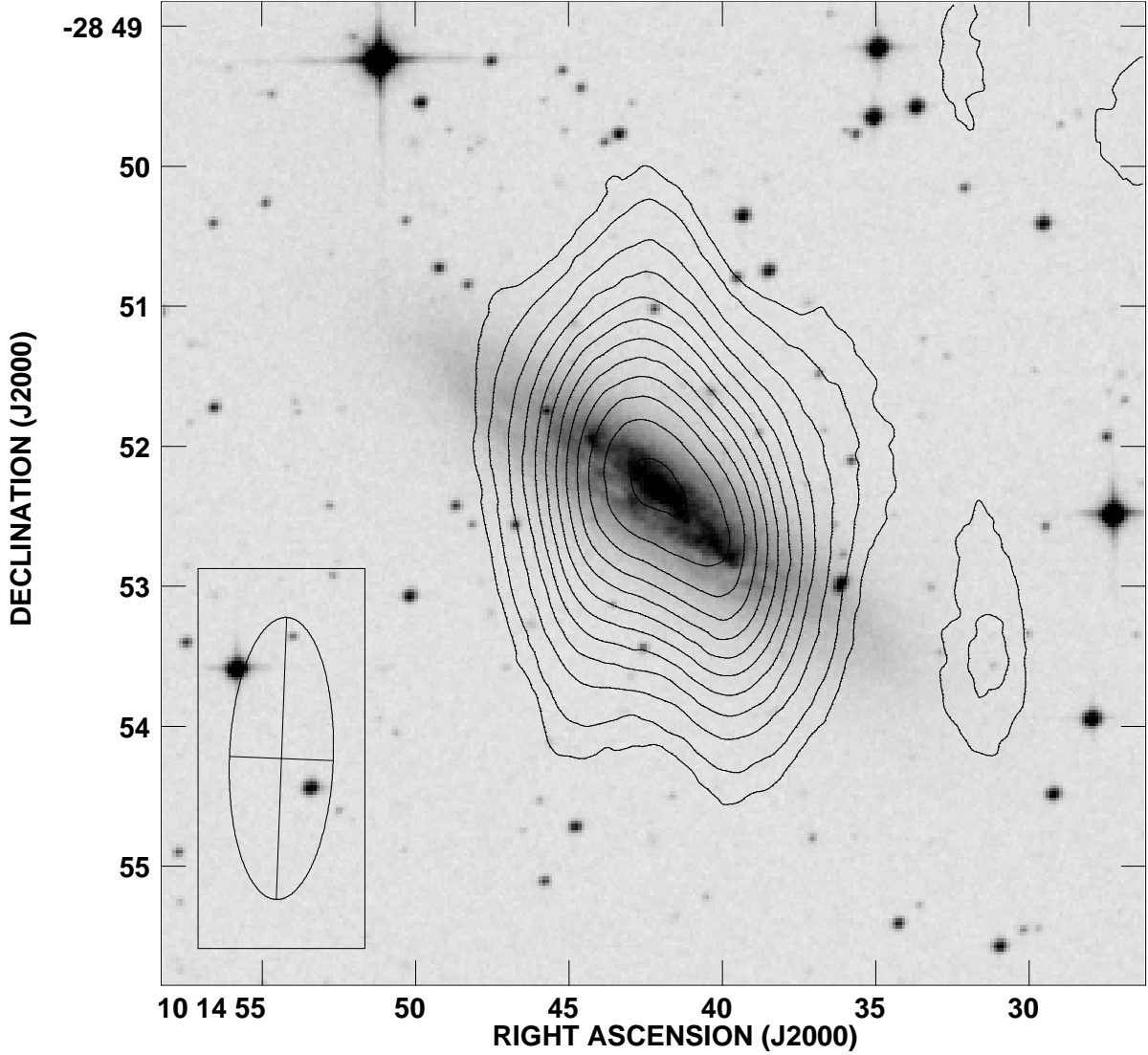}}
\vspace*{-2cm}
\caption{\hi\ total intensity map of NGC\,3175, with a resolution
({\it FWHM}) of $121''\times44\farcs5$ (see beam profile in the 
lower left), overlaid on a Digital Sky Survey image. The contour 
levels are 0.42, 0.6, 0.9, 1.2, 1.5, ... 3.6 Jy beam$^{-1}$ km 
s$^{-1}$.
\label{fig:mom0}}
\end{figure*}

A large number of ``typical'' global properties of galaxies of 
various Hubble types are presented and discussed by Roberts \&
Haynes (1994; in the following RH94). We will use these as 
standard reference values, with respect to which we will discuss 
our results from \hi\ observations of the southern edge-on spiral 
galaxy NGC\,3175. 
RH94, as well as e.g. Rhee \& van Albada (1996), determined that 
typical $L_*$ spiral galaxies, i.e. galaxies near the ``knee'' of 
the galaxy luminosity function, with an absolute $B$ magnitude of 
$M_{\rm B}\ \simeq\ -20$, normally have \hi\ gas masses on the 
order of $10^9-10^{10}\ M_\odot$. This usually represents a few 
percent of their total mass (RH94). The spatial distribution of 
\hi\ gas is usually more extended than that of the stellar disk 
(Bosma 1981). 

With respect to its global properties the edge-on spiral galaxy 
NGC\,3175 is quite peculiar. Its Hubble type is uncertain; the 
Third Reference Catalogue of Bright Galaxies (RC3; de Vaucouleurs 
et al. 1991) classifies it as SAB(s)a?, while it is listed as Sc 
in the Revised Shapley-Ames Catalog (RSA; Sandage \& Tammann 
1981). In the Uppsala General Catalog of Galaxies (UGC), Nilson 
(1973) claims that de Vaucouleurs' classification (which had
been taken over from the RC2) is incorrect and that NGC\,3175 
is ``probably a late-type galaxy''. Because of this confusion, 
we will compare its global properties with those of galaxies 
ranging from type Sa to Sc.

Although at a redshift of only 1098 km s$^{-1}$ (and thus at 
a distance of only 15.9 Mpc)\footnote{We adopt here $H_0 = 75$ 
km s$^{-1}$ Mpc$^{-1}$ and a virgocentric infall velocity of 
300 km s$^{-1}$.}, this galaxy was, until a few years ago, not 
detected in \hi\ emission (Mathewson et al. 1992).
Recently, Mathewson \& Ford (1996) and Theureau et al. (1998) 
found weak \hi\ emission from NGC\,3175. 
Despite this difficulty in tracing its \hi\ gas, it has been 
readily detected in H$\alpha$ (Ryder \& Dopita 1993, 1994), 
radio continuum (Condon et al. 1996) and CO emission (Elfhag 
et al. 1996), which are tracers of star forming regions. 
The Infra-Red Astronomical Satellite (IRAS) measured 60 $\mu$m 
and 100 $\mu$m fluxes of 13.1 and 28.2 Jy, respectively. With 
an $f_{60}/f_{100}$ flux ratio of 0.46, tracing warm dust, and 
in the absence of a luminous AGN, this galaxy is obviously 
forming stars that heat the interstellar medium (ISM) and dust 
in its disk. 
The detection of CO emission in the central part of NGC\,3175
proves that there is molecular gas from which stars can form.
On the other hand, with only barely detectable \hi\ emission, 
it is not clear where NGC\,3715 has its gas reservoir from
which to form stars in the future. We investigate this riddle 
here based on new \hi\ observations conducted with the Australia 
Telescope Compact Array (ATCA)\footnote{The Australia Telescope 
is funded by the Commonwealth of Australia for operation as a 
National Facility managed by CSIRO.}.

\section{Observations and data reduction} 

Our ATCA observations were obtained on 1997, Jan. 27--28, with 
the 750D configuration over a timerange of 13 hours. The total 
on-source integration time is 10 h. The ATCA correlator setup 
used by us provides a velocity resolution of 3.3 km s$^{-1}$.

1934-638 was used as the primary flux calibrator and 1012-44 as 
phase calibrator. The measured flux of 1934-638 was 14.94 Jy 
at 1.42 GHz. The data reduction was performed in the standard 
fashion, using the software package MIRIAD. In order to 
achieve maximum sensitivity for extended emission, natural 
weighting of the visibilities was used. The disadvantage of 
this procedure, namely the low angular resolution of the 
images, is secondary for the current study.

With an angular extent of its radio emission of only a few 
arcmin (see below), NGC\,3175 fits easily into the primary 
beam of the ATCA's 22 m antennae of $34'$. Therefore, no 
primary beam correction is necessary. With a shortest spacing 
of 31 m, flux losses due to missing short spacings are 
negligible.

\section{Results}

\subsection{H\,I distribution}

Figure~\ref{fig:mom0}\ displays our ATCA image of the total \hi\ 
emission from NGC\,3175, superposed as contours on a Digital Sky 
Survey (DSS) image. The angular resolution ({\it FWHM}) of the 
\hi\ image is $121''\times44\farcs5$, as indicated in the lower 
left. The emission is marginally resolved along the disk plane 
of the galaxy in what appears to be a double-peaked distribution. 

\begin{figure}[t!]
\resizebox{1.0\hsize}{!}{\includegraphics{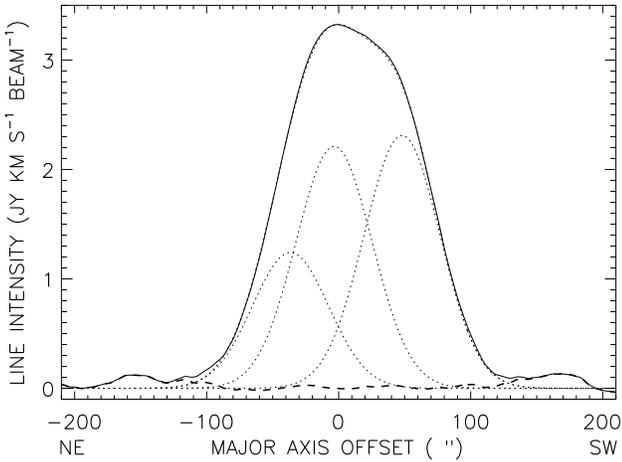}}
\caption{Cut through the \hi\ line emission distribution of NGC\,3175 
along its major axis. The width of each gaussian (dotted lines) is 
that of the angular resolution in the direction of the major axis, 
{\it FWHM} = $67\farcs 7$. Also displayed are the sum of the three
gaussians (dotted line) and the residual after subtracting this
sum from the data (bold dashed line). Measured values are listed
in Table~\protect\ref{tab:rcut}.
\label{fig:rcut}}
\end{figure}

A cut through the \hi\ emission distribution along the galaxy's
major axis, for which we adopt a position angle of {\it PA} $= 
51^\circ$ (Dahlem et al. 2001), is displayed in Fig.~\ref{fig:rcut}.
In this figure the data are represented by a solid line. 

\begin{table*}[t!]
\begin{flushleft}
\leavevmode
\caption{Positions and intensities of emission components along 
  the major axis of NGC\,3175}
\label{tab:rcut}
\begin{tabular}{lcccccc}
\noalign{\hrule\smallskip}
 Component & \multispan{3}{\hspace*{16mm}\hi\,\,Emission\,\,Line$^{\rm a}$} 
  & \multispan{3}{\hspace*{16mm}1.49\,\,GHz\,\,Continuum$^{\rm b}$} \\
 & \multispan{2}{\hspace*{13mm}Radial\,\,Offset} & Relative$^{\rm c}$ & 
   \multispan{2}{\hspace*{13mm}Radial\,\,Offset} & Relative$^{\rm c}$ \\
 & (~$''$) & (kpc) & Intensity & (~$''$) & (kpc) & Intensity \\
\noalign{\hrule\smallskip}
NE     &   $-37.0\pm 3.0$ & $-2.86\pm 0.23$ & $0.56\pm 0.03$ 
  & $-16.7\pm 0.5$ & $-1.29\pm 0.04$ & $0.49\pm 0.02$ \\
Centre & ~\,$-4.0\pm 3.0$ & $-0.32\pm 0.23$ & $1.00\pm 0.02$ 
  & \,~~~$0.0\pm 0.5$ & ~~$0.00\pm 0.04$ & $1.00\pm 0.01$ \\
SW     &   $+47.5\pm 3.0$ & $+3.66\pm 0.23$ & $1.05\pm 0.02$ 
  & $+14.5\pm 0.5$ & $+1.12\pm 0.04$ & $0.34\pm 0.03$ \\
\noalign{\smallskip\hrule}
\end{tabular}
\end{flushleft}
Notes to Table~\protect\ref{tab:rcut}: \\
a) Our data. \\
b) Condon et al. (1996). \\
c) The intensity of the central peak was normalised to unity. \\
\end{table*}

\begin{figure}[t!]
\resizebox{1.0\hsize}{!}{\includegraphics{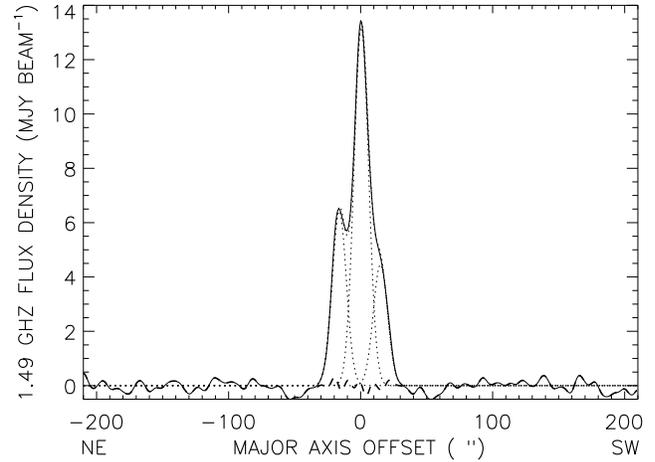}}
\caption{Cut through the 1.49 GHz radio continuum distribution 
of NGC\,3175 along its major axis. The angular resolution is 
$15''$ (Condon et al. 1996). Both the orientation and the 
range of this cut are identical to that presented above in 
Fig.~\protect\ref{fig:rcut}. 
The same three-gaussian approximation is displayed as above.
Measured values are listed in Table~\protect\ref{tab:rcut}.
\label{fig:rcrcut}}
\end{figure}

It became clear very soon during our spatial analysis of the
emission distribution that a two-component approximation is
not adequate. Thus, a three-gaussian model was computed and 
graphically displayed. The individual gaussians and the sum of 
the three are represented by dotted lines. The remaining residual
is shown as a bold dashed line. The width of $67\farcs 7$ of all 
three gaussians is identical; it represents the resolution of our
data in the direction of the major axis over the width of the
cut of $35''$. The fact that the three gaussians leave virtually
no residuals indicates that most of the \hi\ line emission arises 
from three maxima, with almost no emission from further out in 
the disk.
The positions of the three gaussian components along the major
axis and their relative intensities, normalising that of the
strongest peak to unity, are tabulated in Table~\ref{tab:rcut}.
The zero point of the radial axis is the position of the radio
continuum maximum, which we assume to be associated with the
galaxy centre.
A justification for using a three-gaussian approximation lies
in the structure of the 1.49 GHz radio continuum emission
distribution, where three emission maxima are visible (Condon 
et al. 1996). A cut through the 1.49 GHz map (with an angular
resolution of $15''$) is displayed in Fig.~\ref{fig:rcrcut}.
The corresponding offset and intensity values were measured 
by us and are also listed in Table~\ref{tab:rcut}. A direct 
comparison of the radial offsets and flux densities shows that, 
although both emission distributions can be approximated by the 
same model, neither the positions nor the relative intensities 
of the peaks coincide. 
The central \hi\ component is located, within the measuring
accuracy, at the position of the central radio continuum 
peak (and thus the centre of the galaxy). The two outer
\hi\ components are located further out along the major
axis than the extranuclear radio continuum peaks. Star 
formation thus takes place within the \hi\ gas layer in 
the central disk, as might be expected. 

The secondary \hi\ emission feature near the southwestern 
edge of the galaxy disk might be a small companion or a 
weak remnant of a tidal spur or arm. Its recession velocity 
of about 1050 km s$^{-1}$\ does not match up with the 
velocities on that side of the galaxy disk closest to it 
(see below).

\subsection{H\,I kinematics}

\begin{figure}[b!]
\resizebox{1.0\hsize}{!}{\includegraphics{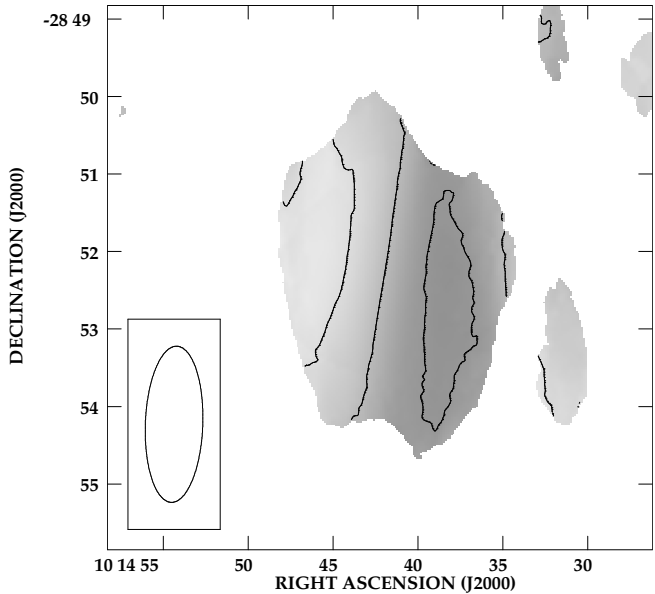}}
\caption{The \hi\ velocity field of NGC\,3175. The grey scale
ranges from minimum (920 km s$^{-1}$; light grey) to maximum 
(1270 km s$^{-1}$; dark grey). The contours shown are 1025 km 
s$^{-1}$, 1100 km s$^{-1}$ (approximately the systemic velocity) 
and 1175 km s$^{-1}$.
\label{fig:vfield}}
\end{figure}

The velocity field (Fig.~\ref{fig:vfield}) shows that the 
lowest velocities are observed in the northeastern part of 
NGC\,3175 and the highest accordingly in the southwestern 
half.\footnote{Based on this knowledge and the convention to
define the {\it PA} as the angle from North to the receding
side of a galaxy, the formally correct value is {\it PA} =
$231^\circ$.}
The dust lane is located on the south-east side of NGC\,3175, 
tracing its ``near side''. This determines the sense of the 
galaxy's rotation uniquely.
The velocity of the secondary emission blob to the southwest
is clearly similar to that of the opposite side of the disk.

\begin{figure}[b!]
\resizebox{1.0\hsize}{!}{\includegraphics{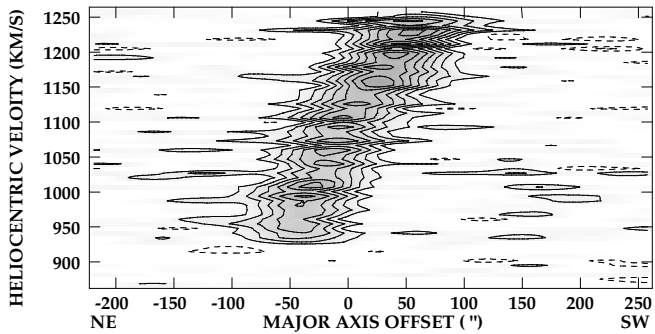}}
\caption{Position-velocity cut through the \hi\ line emission 
distribution of NGC\,3175 along its major axis. The angular 
resolution in the direction of the major axis is $67\farcs 7$, 
the velocity resolution is 6.6 km s$^{-1}$.
\label{fig:pv}}
\end{figure}

A position-velocity (pv) diagram along the major axis of 
NGC\,3175 (along {\it PA} = $231^\circ$) is displayed in 
Fig.~\ref{fig:pv}. One can discern solid body rotation 
out to a radius of about $\pm 45''$ from the centre of the 
galaxy. 
A comparison with Fig.~3 by Mathewson et al. (1992) reveals 
that the velocity gradient is not so low because of beam 
smearing effects, but the same gradient is observed in 
H$\alpha$, with much higher angular resolution. 
However, it is unusual for spirals of types earlier than Scd 
not to have any \hi\ gas beyond the turnover radius in the
rotation curve.

The secondary emission blob is not visible here because 
of its low signal-to-noise ratio. It arises from several
marginally positive signals in the channels in the range
1000--1050 km s$^{-1}$ (at a radial offset of about 
$120''-200''$).

\subsection{Total H\,I line flux and gas mass}

\begin{figure}[t!]
\resizebox{1.2\hsize}{!}{\rotatebox{-90}{\includegraphics{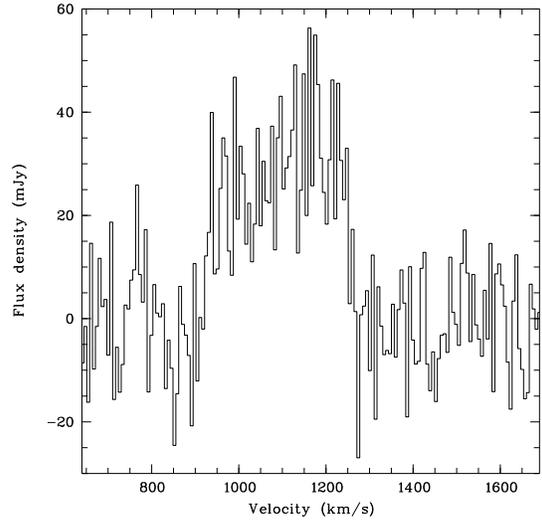}}}
\caption{Integral \hi\ line spectrum of NGC\,3175 measured with
the ATCA. The velocity resolution of the displayed data is 6.6 km 
s$^{-1}$. The velocities on the x-axis are heliocentric.
\label{fig:intspec}}
\end{figure}

The total \hi\ line spectrum of NGC\,3175 is displayed in 
Fig.~\ref{fig:intspec}. The integral \hi\ line spans a 
velocity range of 350 km s$^{-1}$, from 920 to 1270 km 
s$^{-1}$ (cf. also Fig.~\ref{fig:pv}), with an approximate 
width at 20\% of the peak of $W_{20}$ = 330 km s$^{-1}$. 
From the integral spectrum we derive a heliocentric 
systemic velocity of $v_{\rm hel}\ = 1095\pm 10$ km s$^{-1}$,
which is compatible with earlier measurements.
The integral line flux from NGC\,3175 is 11.0 Jy km 
s$^{-1}$, with an estimated uncertainty of about 20\%, 
which is roughly compatible with the non-detection 
reported earlier, with an upper limit of 8.3 Jy km s$^{-1}$\ 
(Mathewson et al. 1992). 
Our \hi\ flux measurement is only slightly lower than the 
values obtained by Mathewson \& Ford (1996) of 14.64 Jy 
km s$^{-1}$ and Theureau et al. (1998) of $12.9\pm1.2$ Jy 
km s$^{-1}$. 
One can estimate the amount of flux that might possibly have
been missed by our interferometer observations by taking the 
flux values from our ATCA data and the one from the Parkes 
data by Mathewson \& Ford (1996) at face value. 
Our measurement reflects the amount of \hi\ gas in the disk 
of NGC\,3175, while the value by Mathewson and Ford might 
be considered as the possible ``full flux'', including low 
surface brightness or intergalactic \hi. One can then argue 
that {\it if} any extended flux had been missed by the 
interferometer, it would not exceed the difference between 
both flux measurements, i.e. about one third of the flux 
displayed in Fig.~\ref{fig:mom0}. 

The 1.344 GHz continuum flux density of NGC\,3175 from our 
ATCA run of $71.5\pm5$ mJy is in very good agreement with 
the value from the VLA data at 1.425 GHz (Condon et al. 1996) 
of 71.8 mJy. Thus, the relative calibration between our data 
and the VLA is good. Because of the weak continuum emission, 
one can assume that the measured total \hi\ line flux is not 
measurably influenced by intrinsic self-absorption.
Our continuum map is not displayed here, because it does not 
add anything new to what is already known from the one by 
Condon et al. (1996).

Using the relation by Roberts (1975) in Eq.~1 and assuming 
optically thin emission, we can calculate the \hi\ gas mass 
in NGC\,3175 from the integral \hi\ line flux as follows:

\begin{equation}
M_{\rm H\,I} = 2.356\,10^5\ D^2\ f_{\rm H\,I}\ [M_\odot] \quad ,
\end{equation}

where $D$ is the distance in units Mpc and $f_{\rm H\,I}$ is 
the measured integral \hi\ line flux. $f_{\rm H\,I}$ = 14.64 
Jy km s$^{-1}$ (from the Parkes data by Mathewson \& Ford 
1996) then translates into a total \hi\ gas mass of NGC\,3175 
of $7.8\,10^8$ \msol, which is quite low for an Sa--Sc type 
spiral. Other late-type spirals have \hi\ gas masses of order 
$10^9-10^{10}$ \msol, typically (e.g. Rhee \& van Albada 1996). 

The \hi\ line flux of $f_{\rm H\,I}$ = 11.0 Jy km s$^{-1}$ 
measured from our data corresponds to an \hi\ gas mass of 
$5.8\,10^8$ \msol. This implies that there are, if any, only 
small amounts ($2\,10^8$ \msol) of intergalactic \hi\ gas in 
the vicinity of NGC\,3175 that might have such a low surface 
brightness as to be missed by the ATCA.

Thus, our present observations prove the presence of small 
amounts of \hi\ gas in the inner disk of NGC\,3175, an area 
where other tracers of star formation (SF) processes had been 
detected previously. Elfhag et al. (1996) report the detection 
of CO emission from the central part of NGC\,3175. It appears 
that the gas from which stars can be formed has finally been 
detected in emission. 

One can use the CO(1--0) line flux derived by Elfhag et al. 
(1996) of $f_{\rm CO}\ = 19.3\pm1.4$ K km s$^{-1}$ (on the
$T_{\rm mb}$ main beam temperature scale) to calculate an
estimate of the total molecular gas mass of NGC\,3175.
This estimate is a lower limit, because only one beam area
({\it FWHM} = $44''$; Elfhag et al. 1996) was observed.
Assuming that the radial CO emission distribution is similar 
to that of the radio continuum, as observed in many galaxies
(see for example Garc\' \i a-Burillo et al. 1992), the CO
emission of NGC\,3175 should arise from within the central
$90''$. At the high inclination angle of NGC\,3175, the
apparent thickness of the disk will probably be only a 
few arcseconds, thus contained within the SEST beam.
Based on the fact that the CO spectrum is peaked near the
systemic velocity of NGC\,3175 it is also likely that the
CO gas distribution is centrally peaked. Therefore, we
estimate that the SEST has gathered {\it at least} 50\% 
of the total CO(1--0) line flux from this object.

We calculate the molecular gas mass, $M_{\rm H_2}$, following
the relation

\begin{equation}
M_{\rm H_2} = I_{\rm CO}\ N({\rm H_2})/I_{\rm CO}\ D^2\ \theta\ 
  m({\rm H_2})\ [M_\odot] \quad ,
\end{equation}

where $I_{\rm CO}\ = 19.3$ K km s$^{-1}$ is the observed 
surface brightness of CO line emission, $N({\rm H_2})/I_{\rm 
CO}$ is the CO-to-H$_2$ conversion factor (``X'' factor) 
between H$_2$ column density and observed CO line surface 
brightness, for which we adopt the ``standard'' value of 
$2\,10^{20}$ cm$^{-2}$/K km s$^{-1}$, $D$ is the distance 
of 15.9 Mpc in units cm, $\theta$ is the area over which 
emission is observed (in units sterad; here the beam {\it 
FWHM} of the SEST) and $m({\rm H_2})$ is the mass of an 
H$_2$ molecule of 1.68\,10$^{-57}$ \msol.

Filling in these quantities, we determine that
$M_{\rm H_2}\ \geq\ 5.8\,10^{8}$ \msol. 
This infers a ratio of $M_{\rm H_2}/M_{\rm HI} \geq\ 0.74$.

In the context of the results by Young \& Knezek (1989),
this makes it likely that NGC\,3175 is an Sc type spiral.
If up to 50\% of the total CO line flux should have been
missed, the corrected ratio is also still consistent with
an Sb--Sbc classification. This ratio of 0.74 would be
unusual for a galaxy of type earlier than Sb though.

\section{Discussion}

\subsection{Global properties of NGC\,3175}

\begin{table}[t!]
\begin{flushleft}
\leavevmode
\caption{Basic properties of NGC\,3175$^{\rm a}$}
\label{tab:basics}
\begin{tabular}{lcc}
\noalign{\hrule\smallskip}
Property & Symbol & Measurement \\
\noalign{\hrule\smallskip}
Absolute blue magnitude        & $M_{\rm B}$ (mag) & $-19.55\pm 0.10$ \\
Total blue luminosity          & $L_{\rm B}$ ($L_\odot$)  
  & $1.03(^{+0.25}_{-0.20})\,10^{10}$ \\
Total \hi\ gas mass       & $M_{\rm HI}$ ($M_\odot$) & $7.8(\pm 1.5)\,10^8$ \\
Total virial mass         & $M_{\rm T}$ ($M_\odot$)  & $7.34(\pm 
  0.85)\,10^{10}$ \\
\hi\ to total mass ratio & $M_{\rm HI}/M_{\rm T}$ & $1.06\%\pm 0.25\%$ \\
\hi\ to blue light ratio & ${M_{\rm HI}\over L_{\rm B}}$ 
  (${M_\odot \over L_\odot}$) & $0.076\pm 0.019$ \\
Total mass-to-light ratio & ${M_{\rm T}\over L_{\rm B}}$ 
(${M_\odot \over L_\odot}$) & $7.13^{+1.78}_{-1.98}$ \\
Optical colour index$^{\rm b}$ & {\it B--V} (mag) & $0.90\pm 0.14$ \\
Total FIR luminosity & $L_{\rm FIR}$ ($L_\odot$) & $6.38(\pm 0.6)\,10^{9}$\\
%
\noalign{\smallskip\hrule}
\end{tabular}
\end{flushleft}
Notes to Table~\protect\ref{tab:basics}: \\
a) All values calculated by us based on $D = 15.9$ Mpc. \\
b) Both $B$ and $V$ magnitudes are from the RC3. \\
\end{table}

In order to interpret the \hi\ gas properties of NGC\,3175 in
general terms and compare the galaxy with other spirals, we
have calculated a few standard properties, as tabulated in
Table~\ref{tab:basics}. The far-infrared luminosity, $L_{\rm 
FIR}$, was taken from Dahlem et al. (2001).

The total virial mass, $M_{\rm T}$, was calculated using the 
formula

\begin{equation}
M_{\rm T}(r) = 2.33\,10^5\ r\ v^2(r)\ [M_\odot] \quad ,
\end{equation}

where $r_{25}\ = 2\farcm 5$, the 25th magnitude isophotal radius, 
was used as a measure of $r$ and half the width of the integral 
\hi\ spectrum, $W_{20}/2 = v_{\rm max}\ = 165$ km s$^{-1}$, as 
$v(r)$ (see Figs.~\ref{fig:pv}\ and \ref{fig:intspec}), adopting
an inclination of $i \simeq\ 90^\circ$. 

From a comparison of the values in Table~\ref{tab:basics}\ with
the data by RH94 (in particular their Figs.~2--5), one can
conclude the following:

\begin{itemize}

\item The total blue luminosity, $L_{\rm B}$, of NGC\,3175 is
	quite normal for a spiral of type Sa--Sc.

\item Its total virial mass, $M_{\rm T}$, lies near the
	25\%-level for galaxies of type Sa--Sc in the sample
	by RH94.

\item Accordingly, the total mass-to-light ratio of NGC\,3175, 
	$M_{\rm T}/L_{\rm B}$, also lies near the 25\%-level in
	RH94's sample.

\item The total \hi\ gas mass of NGC\,3175 is relatively low
	(near the 25\%-level) if it were an Sa or Sab galaxy; 
	for types Sb--Sc, it is well below the 25th percentile
	mark of the data used by RH94, indicating an \hi\ 
	deficiency. 

\item The ratio of \hi\ gas mass to total mass of only about
	1\% is at about the 25\%-level in RH94's data
	if it were an Sa or Sab galaxy. For all later types, 
	this percentage represents a pronounced deficiency
	(because there is little variance in the ratios for
	spirals of type Sb or later). Note that if we use 
	our own \hi\ line flux measurement this deficiency 
	is yet much more pronounced, with an \hi -to-total
	mass ratio of only 0.8\%.

\item The same is true for the $M_{\rm HI}/L_{\rm B}$ ratio.

\item Even for an Sa galaxy, NGC\,3175's {\it B--V} colour is
	unusually red, being far beyond the 75\%-level of
	RH94's data. For galaxies of later types the
	discrepancy is yet more pronounced. NGC\,3175
	is in fact one of the reddest nearby spiral galaxies,
	with a predominantly old stellar population. The
	same was already found by Ryder \& Dopita (1993).

\item On the other hand, its FIR luminosity, $L_{\rm FIR}$, 
	is normal for galaxies of types Sa--Sc.

\end{itemize}

One should note that $L_{\rm B}$, $M_{\rm T}$ and $L_{\rm FIR}$ 
do not show a strong dependence on Hubble type (see RH94). On the 
other hand, quantities like $M_{\rm HI}$, $M_{\rm HI}/L_{\rm B}$
and $M_{\rm HI}/M_{\rm T}$ do show a dependence of Hubble type 
(albeit with considerable overlaps). According to these properties, 
NGC\,3175 shows an \hi\ deficiency. For type Sa--Sab, this 
deficiency would not be very pronounced, but if NGC\,3175 
should be of a later Hubble type, its \hi\ content is clearly 
deficient.

\subsection{Star formation in the central disk}

\hi\ gas has been found by us, corroborating earlier detections;
as stated above, NGC\,3175 is \hi --deficient. The ATCA data 
provide for the first time spatially resolved information.
Our total \hi\ intensity map in Fig.~\ref{fig:mom0}\ indicates 
that most of the detected \hi\ gas is located within the central 
part of the galaxy. Because this is difficult to quantify by
looking at the map only, we have produced the radial cut through
the emission distribution shown in Fig.~\ref{fig:rcut}.
From the position of the gauss components (see also 
Table~\ref{tab:rcut}) it follows that practically all \hi\
gas is contained within the central 7 kpc, and that the ongoing 
SF is enveloped by \hi\ gas. It is very unusual that \hi\ is 
detected only in the central $90''$ (7 kpc) of NGC\,3175, while 
normally the stellar disks of spiral galaxies are embedded in 
extended \hi\ envelopes (Bosma 1981). 
The radial extent of the \hi\ distribution in NGC\,3175 roughly 
coincides with that of the H$\alpha$ and 1.4 GHz radio continuum 
emission (Ryder \& Dopita 1993; Condon et al. 1996). Therefore, 
only the central part of the disk is apparently forming stars at 
the present time. This is corroborated by the fact that this is
the only region in NGC\,3175 with optical colours suggesting the
presence of a young population of stars (Ryder \& Dopita 1994).
All over this central area the \hi\ gas follows a solid body 
rotation pattern (Fig.~\ref{fig:pv}).

There is no reliable estimate on the total molecular gas content 
of NGC\,3175 in the literature and it is doubtful that Elfhag et 
al. (1996) measured the total CO line flux of NGC\,3175 within 
one $44''$ beam of the Swedish-ESO Submillimetre Telescope. 
Still, the CO spectrum by Elfhag et al. (1996) can serve to give 
a first impression of the molecular gas properties of the nuclear 
area in NGC\,3175. Its peak antenna temperature of ca. 0.15 K is 
quite typical for galaxies at a distance of 15.9 Mpc (see also
for comparison Young et al. 1995). This and the numbers derived
above suggest that \hi\ gas, which is normally more extended than 
CO, might be more deficient than molecular gas. 

Inside the inner disk of NGC\,3175, within a radius of 3.5 kpc, 
the density of cold atomic and molecular gas appears to be high 
enough for the onset of SF. 
In fact, the FIR properties of the inner disk in NGC\,3175, 
especially its $f_{60}/f_{100}$ flux ratio of 0.46 and the 
implied warm dust temperatures, are reminiscent of starburst 
galaxies with disk-halo interactions (e.g. Heckman et al. 1990; 
Dahlem 1997). 
This is consistent with the observed radio continuum emission
distribution which indicates the presence of an outflow from 
the \hii\ regions in the central disk into the halo (Condon et 
al. 1996, Dahlem et al. 2001). Since all SF activity is visibly 
occurring in the central 7 kpc of NGC\,3175, it can be assumed 
that the FIR emission comes from that same region.

\subsection{The quiescent outer disk of NGC\,3175}

Virtually no \hi\ gas was detected beyond a radius of 3.5 kpc
(see Fig.~\ref{fig:rcut}) with the sensitivity of the current
data. At the same time, the outer disk of NGC\,3175 exhibits 
no measurable signs of ongoing SF. There is neither H$\alpha$ 
nor radio continuum emission and there is also no sign of a 
young, blue population of stars. The outer disk of NGC\,3175 
does show signs of spiral arms; thus it can be assumed to 
rotate differentially, but there is not enough gas to trace 
its rotation pattern.

Such a situation, and also the observed activity in the inner
disk, might be created by an interaction (possibly a distant 
passage of a companion) in the past, which could have affected 
the outer, volatile neutral atomic gas component, but not the 
more centrally concentrated and therefore more stably bound 
molecular gas (e.g. Kenney \& Young 1989). RH94 and Giovanelli 
\& Haynes (1988) note that interacting galaxies in groups can 
lose up to 50\% of their gas, galaxies in clusters even up to 
90\%. 

All this would also be consistent with the exceptionally red 
colour of NGC\,3175's disk beyond a radius of 3.5 kpc, because 
a gravitational interaction could remove part of the gas from 
the outer disk, while a fraction of the remaining gas would 
move into the central region (e.g. Combes 1987). 
The outer, quiescent disk dominates the mean {\it B-V} colour 
index of NGC\,3175, because it covers about 90\% of the surface 
area.

\subsection{The ``missing'' H\,I gas}

Our \hi\ observations cannot explain where the \hi\ gas is that
must at one point in time have been present in the outer disk
of NGC\,3175. 
Except the one weak emission blob to the south-west of NGC\,3175, 
which is negligible in terms of its gas mass (0.8 Jy km s$^{-1}\ 
\hat =\ 4.2\,10^6$ \msol), our \hi\ data show no sign of emission 
from either tidal features (like tails, plumes, tidal arms or any 
other sort of intergalactic or galactic emission from a potential 
optically faint partner) within a radius of $15'$ (70 kpc). 
These results are consistent with those of Garc\'\i a (1993), 
according to whom NGC\,3175 is a member of a group of 5 galaxies, 
none of which is close enough (in spatial projection and at the
same time in velocity space) to be directly interacting with it 
at the current time (Table~2 by Garc\'\i a 1993). 

A look at the DSS does not help in proposing a likely scenario 
for the removal of gas from NGC\,3175 either. There is no 
visible interaction partner near NGC\,3175 (within a projected 
distance of $50'$, i.e. 230 kpc) massive enough to have stripped 
it of a significant fraction of its \hi\ gas. ESO\,436-G004, 
about $5'$ south of NGC\,3175, is at a different redshift and
thus distance (and therefore not detected in our observations). 
%
%
Two small objects $1'$ west of ESO\,436-G004 that are visible 
on the second generation DSS (DSS-2) plates, if at approximately
the distance as NGC\,3175, must be dwarf galaxies.

Our ATCA data are most sensitive to structures with a spatial 
extent similar to the beam size and the sensitivity after only 
one aperture synthesis is still limited. Thus, a non-detection 
of intergalactic \hi\ gas in this data set does not necessarily 
impose strong constraints on its potential properties. 
However, the presence of substantial amounts of \hi\ gas in an 
area of about $10'$ around NGC\,3175 is ruled out by the Parkes 
data (see above). Also the new Parkes multi-beam survey database, 
which was searched by us, does not exhibit any signs of \hi\ 
emission in the relevant redshift interval closer to NGC\,3175 
than $1\fdg 3$ (or 360 kpc), where NGC\,3137, one of the group 
members, is detected. 
Therefore, the riddle about the fate of NGC\,3175's \hi\ gas 
remains unsolved for the time being. It is not clear where it 
might have gone or whether NGC\,3175 ever had more \hi\ gas 
than is detectable at the present time. Certainly there must 
have been \hi\ gas in the outer disk before from which the 
observed old stellar population formed.

The fact that NGC\,3175 is not a member of a cluster of galaxies, 
but at the same time \hi --deficient, makes this deficiency with
simultaneously ongoing SF in its inner part so 
intriguing. However, a more solid interpretation of this issue 
must await high-resolution \hi\ and CO imaging observations. 

\begin{acknowledgements}
%
We thank the referee, Dr. A. Bosma, for useful comments that
helped improve the paper considerably and Dr. J. Condon for
making available the 1.49 GHz radio continuum map in digital
form.
The Digitized Sky Survey was produced at the Space Telescope 
Science Institue under U.S. Government grant NAG W-2166. The 
National Geographic Society - Palomar Observatory Sky Atlas 
(POSS-I) was made by the California Institute of Technology 
with grants from the National Geographic Society. 
\end{acknowledgements}


\begin{thebibliography}{}

\bibitem[]{}
Bosma A. 1981, AJ 86, 1825

\bibitem[]{}
Combes F. 1987, in Proceedings NATO Conference on ``Galactic
  and Extragalactic Star Formation'', R. Pudritz and M. Fich
  (eds.), Reidel, Dordrecht, p. 475

\bibitem[]{}
Condon J. J., Helou G., Sanders D. B., Soifer B. T. 1996, ApJS 
  103, 81


\bibitem[]{}
Dahlem M. 1997, PASP 109, 1298

\bibitem[]{}
Dahlem M., Lazendic J., Haynes R. F., Ehle M., Lisenfeld U. 2001, 
  A\&A (subm.)


\bibitem[]{}
Elfhag T., Booth R. S., H\"oglund B. et al. 1996, A\&AS 115, 439

\bibitem[]{}
Garc\'\i a A. M. 1993, A\&AS 100, 47

\bibitem[]{}
Garc\'\i a-Burillo S., Gu\'elin M., Cernicharo J. J., Dahlem M.
  1992, A\&A 266, 21

\bibitem[]{}
Giovanelli R., Haynes M. P. 1988, in ``Galactic and Extragalactic
  Radio Astronomy'', G. L. Verschuur, K I. Kellermann (eds.),
  Springer Verlag, Heidelberg, p. 522

\bibitem[]{}
Heckman T. M., Armus L., Miley G. K. 1990, ApJS 74, 833

\bibitem[]{}
Kenney J. D. P., Young J. S. 1989, ApJ 344, 171

\bibitem[]{}
Mathewson D. S., Ford V. L. 1996, ApJS 107, 97

\bibitem[]{}
Mathewson D. S., Ford V. L., Buchhorn M. 1992, ApJS 81, 413


\bibitem[]{}
Nilson P. 1973, ``Uppsala General Catalog of Galaxies'',
  Uppsala Astronomical Observatory 

\bibitem[]{}
Rhee M.-H., Albada T. S. 1996, A\&A 115, 407

\bibitem[]{}
Roberts M. S. 1975, in ``Galaxies and the Universe'', A. Sandage,
  M. Sandage, J. Kristian (eds.), Cambridge University Press,
  Chicago

\bibitem[]{}
Roberts M. S., Haynes M. P. 1994, ARA\&A 32, 115

\bibitem[]{}
Ryder S. D., Dopita M. A. 1993, ApJS 88, 415

\bibitem[]{}
Ryder S. D., Dopita M. A. 1994, ApJ 430, 142

\bibitem[]{}
Sandage A. R., Tammann G. A. 1987, ``A Revised Shapley-Ames
  Catalog of Bright Galaxies'', Carnegie Institution of Washington
  Publ. 635, Washington, 2nd edition


\bibitem[]{}
Theureau G., Bottinelli L., Coudreau-Durand N. et al. 1998, 
  A\&AS 130, 333


\bibitem[]{}
de Vaucouleurs G., de Vaucouleurs A., Corwin H. G., Buta R. J.,
  Paturel G., Fouqu\' e P. 1991, ``Third Reference Catalogue of
  Bright Galaxies'', Springer Verlag, New York

\bibitem[]{}
Young J. S., Xie S., Tacconi L., et al. 1995, ApJS 98, 219
%

\bibitem[]{}
Young J. S., Knezek P. M. 1989, ApJ 347, L55

\end{thebibliography}
\end{document}